# Quick-cast: A method for fast and precise scalable production of fluid-driven elastomeric soft actuators

Bratislav Svetozarevic[1]*, Moritz Begle[1]*, Stefan Caranovic[1], Zoltan Nagy[2], Arno Schlueter[1]

[1] Architecture and Building Systems, Department of Architecture, ETH Zurich, Switzerland

[2] Intelligent Environments Laboratory, Department of Civil, Architectural and Environmental Engineering, The University of Texas at Austin, TX 78712, USA

*These authors contributed equally to the work.

**Abstract**

Fluid-driven elastomeric actuators (FEAs) are among the most popular actuators in the emerging field of soft robotics. Intrinsically compliant, with continuum of motion, large strokes, little friction, and high power-to-weight ratio, they are very similar to biological muscles, and have enabled new applications in automation, architecture, medicine, and human-robot interaction. To foster future applications of FEAs, in this paper we present a new manufacturing method for fast and precise scalable production of complex FEAs of high quality (leak-free, single-body form, with <0.2 mm precision). The method is based on 3d moulding and supports elastomers with a wide range of viscosity, pot life, and Young's modulus. We developed this process for two different settings: one in laboratory conditions for fast prototyping with 3d printed moulds and using multi-component liquid elastomers, and the other process in an industrial setting with 3d moulds micromachined in metal and applying compression moulding. We demonstrate these methods in fabrication of up to several tens of two-axis, three-chambered soft actuators, with two types of chamber walls – cylindrical and corrugated. The actuators are then applied as motion drivers in kinetic photovoltaic building envelopes.

## 1 Introduction

Soft robotics is a rapidly growing field offering novel actuators [1], end effectors [2], skins [3], and even entire robots [4] made of compliant material. They carry several important advantages compared to classical rigid-bodied robots, such as intrinsic compliance, large power-to-weight ratio, low friction, simpler control, and cheaper fabrication, and have enabled solutions to challenging problems in automation (e.g. universal grasping [5]), architecture (e.g. adaptive solar façade [6]), medicine (e.g. minimally invasive surgery [7]), and human-robot interaction (e.g. safe continuum manipulation [8]). As contrary to rigid-bodied robots, where components are selected from the standard set of components (e.g. electromotor, gearbox, spring) and designs follow rigid-body kinematics principles (rigid links connected by discrete joints), soft robotics offers much more design freedom, with a plethora of soft materials available (e.g. intrinsically compliant – elastomers, and extrinsically compliant, such as wires of shape memory alloy [9], thermoplastics [10], and jamming particles [5]), a range of actuation principles (fluidic-, thermal-, humidity-, pH-, magnetic-, and electric-driven), and freedom in defining components shape, structure, and working principle. Consequently, the main challenge in working with soft components is in their design and manufacturing. Compared to design, manufacturing techniques are directly linked to the performance of the actuators, which often brought major advances in the past [11]–[14].

Among soft actuators, one of most popular types are fluid-driven elastomeric actuators (FEAs). These actuators contain fluidic pathways within their soft bodies. When they are filled up with pressurised fluid (gas or liquid), the surrounding elastic material expands, leading to a change of the outer shape of the actuator. Depending on the geometry of fluidic pathways and shape of the actuator, as well as on the pressurisation method (inflation or deflation), different output motion paths can be obtained, such as expansion, contraction, bending, or twisting [15]. Besides motion diversity, FEAs exhibit advanced actuation features, similar to those of biological muscles, such as continuum of motion, high power-to-weight ratio, large strokes, and little friction. Also, due to use of elastomers, FEAs are intrinsically compliant and can passively mitigate external disturbances (acting as a spring-damper system), allowing for reduction of control complexity and safe interaction with the environment and humans. For existing reviews of FEAs, please see [12], [16].

Even though several methods for manufacturing of FEAs exists, they are of limited suitability for scalable production due to either high process complexity, long total fabrication time, difficulties in achieving leak-free actuators, or limited support for different types of elastomers. To strengthen future applications of soft robotics (e.g. soft-robotic driven adaptive solar facades [17]), a manufacturing method suitable for scalable production of FEAs is necessary.

In this paper we present a new manufacturing method for soft actuators that is suitable for fast and precise scalable production of complex FEAs of high quality (leak-free, single-body, with <1 mm precision). This method is based on 3d moulding, where outer walls of an FEA and internal voids (fluidic pathways) are formed in a single casting process. The outer mould consists of several parts that can be easily disassembled, while internal voids are created by pulling out the mould parts. In that, the elasticity of elastomers enables pulling out of the three internal mould parts.

The advantages of this method are: (i) speed – a single casting method with the total production time of approximately one hour per actuator, (ii) quality of the actuators – single-body (no gluing or bonding required), leak-free, and precise (<0.2 mm) actuators, and (iii) a wide range of supported elastomers in terms of viscosity, pot life, and Youngs' modulus. We call the method shortly Quick-cast. We developed this process for two different settings: one for fast prototyping in laboratory conditions with 3d printed moulds and using multi-component pourable elastomers, and the other process in an industrial setting with 3d moulds micromachined in metal and using compression moulding. We demonstrate these methods in fabrication of up to several tens of two-axis, three-chambered FEAs, with two types of chamber walls – cylindrical and corrugated. We tested them in terms of pressure-deflection characteristics and motion repeatability. The actuators are then used as motion drivers in two real-world building-scale prototypes of kinetic photovoltaic envelopes, with 50 modules at ETH Zurich Hoengerberg Campus [6] (Fig. 1) and with 30 modules at EMPA, Duebendorf [17], both in Switzerland.

These novel dynamic building envelopes are very lightweight, resilient to weather condition, robust to wind loads, with low self-power consumption, and can improve a building's net energy demand through adaptive shading and electricity generation [17], [18]. Besides the three-chambered actuator design reported here, the Quick-cast method is suitable for other single or multi-chamber designs.

## 2 Methods and materials

*2.1 Actuator design*

We developed the Quick-cast manufacturing method from the need to fabricate a large number (>100) of two-axis three-chambered single-body FEAs, called SoRo-Tracks (Fig. 1h) [19], in order to apply them as motion drivers in adaptive photovoltaic façades [6], [17]. In terms of motion capabilities, we decided for two-axis actuation instead of single-axis. Two-axis allow for both vertical (altitude angle) and horizontal positioning (azimuth angle) of façade elements, which increases the yearly PV energy output by 10-15% compared to single-axis trackers [20]. Furthermore, the possibility to move each of the façade elements in two-axis provides architects with the maximum design freedom when it comes to aesthetic expressions. In terms of actuator design requirements, we wanted to reduce its complexity as much as possible, as well as to improve its visual appearance and allow for easier cleaning by having a single-body actuator, instead of, for example, multiple single-axis soft actuators.

We considered two geometries of internal voids: cylindrical and corrugated. The cylindrical is simpler in terms of design complexity and it was the very first intuitive design to test the method. The corrugated design has more complex geometry, offers much more design parameters (e.g. ribs geometry, walls thicknesses), and therefore requires more effort to obtain a well performing actuator. In the design of bending actuators, it is important to maximise the forces acting on bottom and top discs of the actuator and minimise the radial expansion of rubber. This would maximise the bending moment of the actuator. Also, it is preferable that the actuator stays in the linear region of rubber elasticity during inflation as then it allows for implementation of open-loop (sensorless) control, instead of sensor-based closed loop control. In our case, the ribs are very precisely designed in 3d in such a way that the inner ribs have thinner walls than the outer ones, and therefore they unfold to the outside, increasing the distance between the outer ribs. The outer ribs have a rather high thickness preventing overall radial expansion of the chamber, and therefore maximising the bending moment. In this way, we avoid large expansions of the rubber and the corrugated SoRo-Track actuator functions in the linear region (Fig. 1g).

*2.2 State-of-the-art manufacturing methods*

We defined the following process properties as important for the scalable production in an industrial setting: (i) achieving leak-free actuators, (ii) number of manufacturing steps, (iii) steps complexity, (iv) required gluing / bonding step, (v) total fabrication time, (vi) precision, (vii) supporting wide range of elastomers, and (ix) possibility to base it on an already established industrial process. In the following text we review the state-of-the-art methods in terms of these properties.

At macro scale (cm to m), moulding and direct 3d printing of soft materials are the most popular manufacturing techniques, due to the fact that elastomers can be processed as liquids – multi-component elastomers that harden over time, with the pot life from min to hours. A typical moulding technique is a 2d moulding process (a layup process) where a part of the actuator containing the fluid pathways is moulded first, and then it is glued to another, stiffer elastomer or fabric, forming networks of pneumatic channels, PneuNets [13]. This method is very popular in the community, due to the availability of 3d printers for fabrication of plastic moulds (ABS, PLA, etc). The drawback of this process is in the required gluing or bonding of separately produced elastomer pieces, which might result in weak spots prone to delamination and leaking at higher pressures. Also, the process consists of several steps, requiring several hours for fabrication of a single actuator.

To overcome the drawbacks of the 2d moulding process, a 3d moulding technique using wax cores has been proposed [14]. In this process, first, the wax cores are casted, and then they are used in the second casting step for making fluidic pathways. This process enables fabrication of very complex fluidic pathways in a single-body actuator form (avoiding gluing), with uniform material properties across the actuator. This process is, however, more time consuming than 2d moulding, as it consists of two castings and, in addition, curing (cross-linking of elastomer chains) cannot be made faster by heating up the mould, due to the low melting point of the beeswax of 62-64ºC. Also, the wax needs to be removed and cleaned from the fluidic pathways afterwards, which is an additional step.

Besides moulding techniques, direct 3d printing of soft materials offers advanced possibilities, such as varying elastomer parameters across the actuator volume and achieving higher geometrical complexity [21]–[23]. However, due to difficulties in obtaining cross-links between polymer chains during the layering process, obtaining leak-tight actuators is often challenging. There are multiple parameters of the 3d printing process (e.g. layer height, nozzle size, and extrusion temperature) that need to be properly tuned for each material, and the 3d printed actuators need to be post-processed to achieve polymer cross-linking [24]. Moreover, current soft-material 3d printing techniques have lower precision than moulding processes and fabrication time per single actuator is much longer.

In terms of manufacturing methods suitable for scalable production, only one method has been proposed in the literature to the best of our knowledge. This process is based on a standard industrial process, rotational casting, where uncured elastomer is casted as a hollow structure in a closed mould [25]. The limitation of this process is that it supports elastomers with limited range of viscosity, pot life, and Youngs' modulus (uncured elastomer needs to be of a certain viscosity in order to be processed in rotational casting). Also, the precision of internal voids depends on how good the models of the rotational casting process and elastomer curing process are, which might be additional limiting factor for the process applicability, in particular for achieving complex and precise fluid pathways. Furthermore, it is a multi-step process, where multiple individual balloons (chambers) obtained in the rotational casting step are then used to produce the final actuator by casting the surrounding elastomer body. In that, firm positioning of the individual balloons is required, which typically needs additional engineering effort.

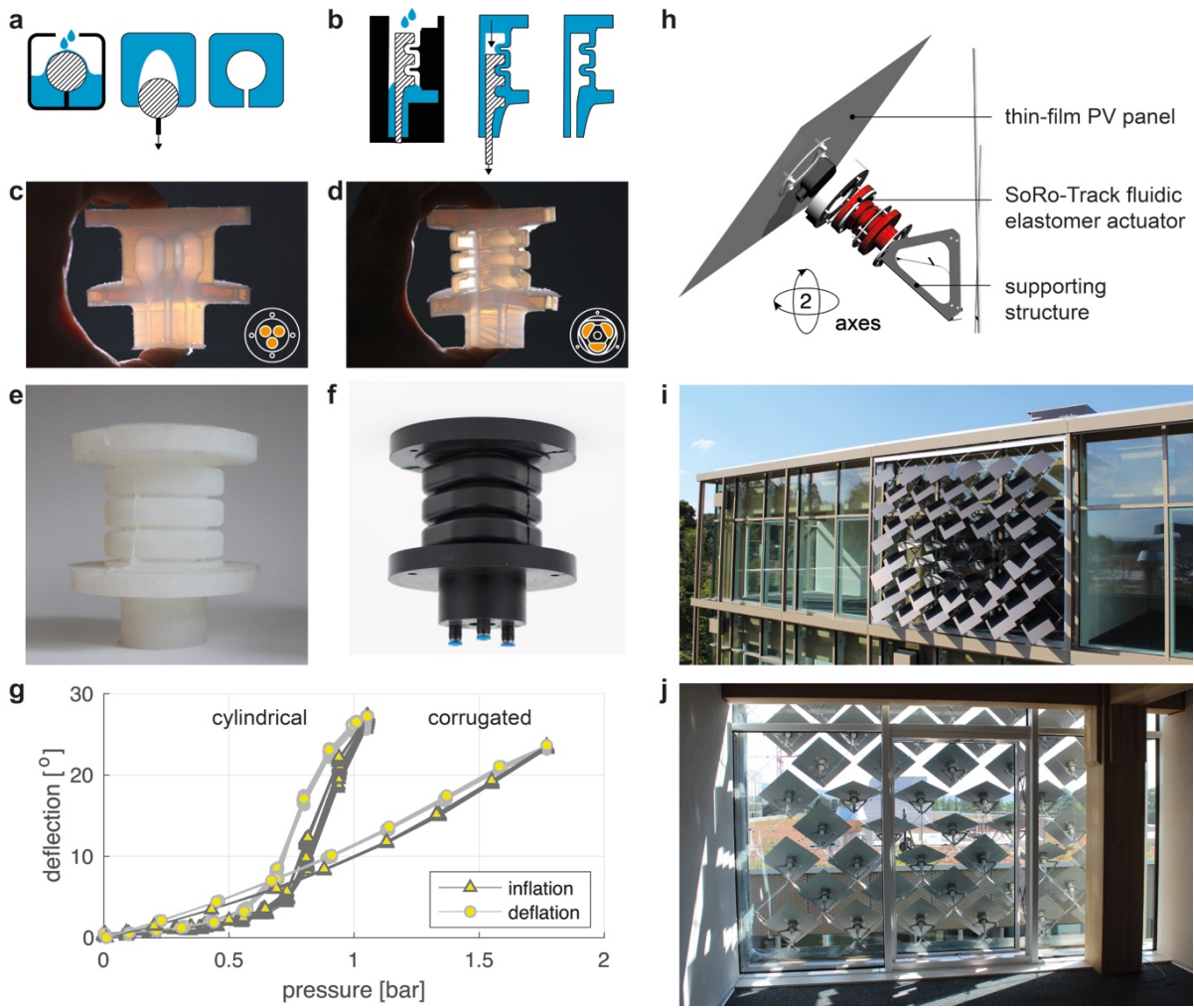

**Fig. 1.** Quick-cast method applied to fabrication of FEAs with cylindrical (a) and corrugated (b) walls. See-through images of cylindrical (c) and corrugated (d) two-axis, three-chambered SoRo-Track actuators made of semi-translucent ELASTOSIL® VARIO from Wacker Chemie AG. Corrugated SoRo-Tracks made of (e) ELASTOSIL® VARIO in laboratory conditions with 3d printed moulds. (f) Corrugated SoRo-Track industrially manufactured from Neoprene rubber. (g) Pressure-deflection characteristics of cylindrical and corrugated SoRo-Tracks. (h) A single module of Adaptive Solar Façade driven by SoRo-Track. Outside view (i) and view from inside (j) of Adaptive Solar Façade with 50 cylindrical SoRo-Tracks at ETH Zurich, Hönggerberg Campus, Switzerland.

## 2.3 Quick-cast manufacturing method in laboratory conditions

We aimed for developing a manufacturing method based on 3d printed moulds as it allows for quick prototyping and testing of soft actuators. After the soft actuator and its mould was modelled in Rhinoceros® (Step A), it was 3d printed from Nylon on a laser sintering system EOS P396 (Step B). The details of the mould are presented in Fig. A.2. After assembling the mould, a multi-component liquid elastomer ELASTOSIL® VARIO from Wacker Chemie AG is poured (step 1). VARIO is a two-component silicone rubber, with Shore A hardness 15 and 40, respectively, which allow for achieving any Shore A hardness between 15 and 40. The influence of the mixing ratio on rubber parameters (tensile strength, elongation at break, and tear strength) are given in Fig. A.1. We mixed the two elastomers in a 3:1 ratio and added 15% of the catalyst (resulting Shore A index 30). After pouring, the mixture was degassed to remove the trapped air-bubbles (Step 2). Then, the curing of elastomer was done in an oven at 75ºC for about 20 min (Step 3). This is much faster than curing at the room temperature, which takes several hours. Finally, the mould is disassembled in two steps (Step 4). First, the two parts forming the outer shape of the actuator are taken apart. Second, the inner cores are pulled out of the soft actuator (Fig. A.3 and A.5). Pulling out of the hard cores is possible due to the large elongation at break of elastomers (in this case it is around 500%; may be up to 1000%, e.g. of Ecoflex™ 00-50 from Smooth-On, Inc.). To allow for easer pulling out of the inner cores, we were slightly inflating the chambers. The total fabrication time of a single actuator, including the mould assembling, takes approximately one hour. To produce the next actuator, one needs to repeat steps 1 to 4. Obtained actuators may be then tested for performance (Step C) and the information fed back to the modelling software to inform the change of the design. The images of this manufacturing process for cylindrical SoRo-Track and using VARIO are given in Fig. A.3. Also, we show the manufacturing of the same actuator with ELASTOSIL® M4601 (Fig. A.4.).

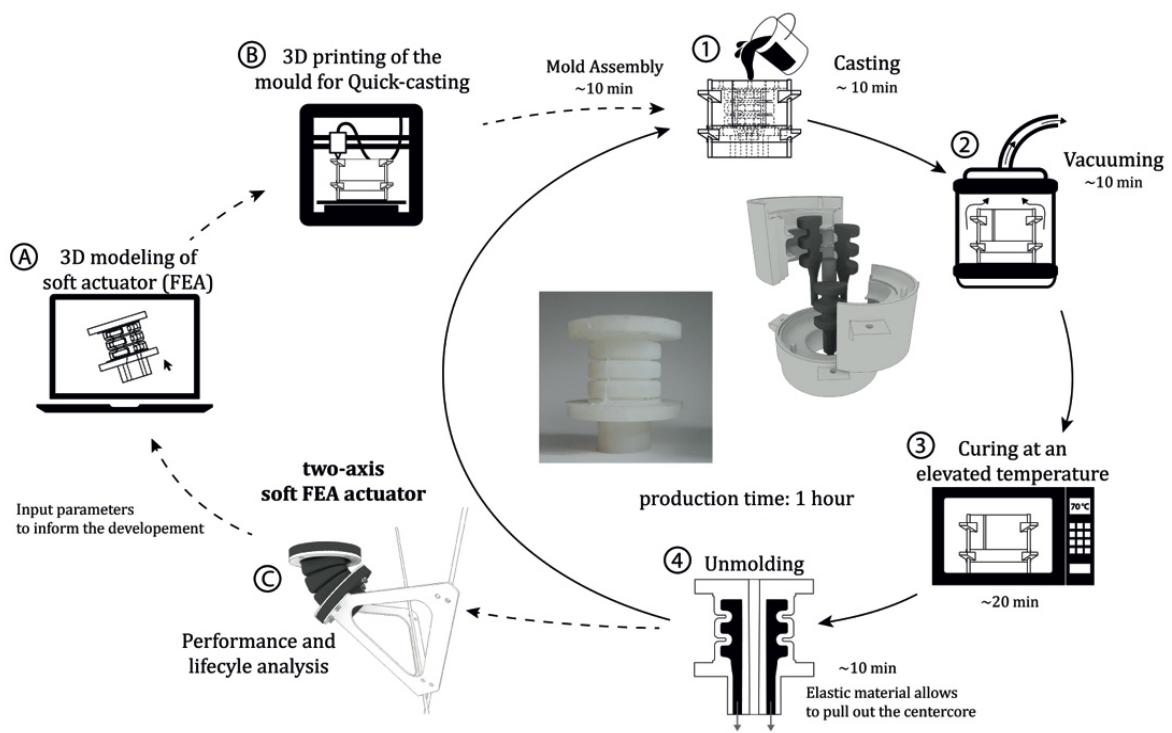

**Fig. 2.** Quick-cast manufacturing method for fast prototyping of FEAs in laboratory conditions with 3d printed moulds and using multi-component liquid elastomers.

*2.4 Quick-cast manufacturing method in industry*

The Quick-cast method in laboratory conditions was adapted to industrial setting, in order to obtain the industry-quality actuators, with repeatable characteristics over a large number of samples. We found compression moulding to be a suitable process, which allows for both multi-component elastomers (e.g. ELASTOSIL® M4601 and ELASTOSIL® VARIO 15/40) and rubbers (e.g. Neoprene and EPDM) to be used. We only tested this method with rubbers. Because of that, the mould is heated up to 350ºC and the uncured rubber is injected under pressure into the metal mould (Step 1). After curing at this elevated temperature for 20 minutes (Step 2), the mould is cooled down and taken apart (Step 3). The disassembling of the mould happens in the same way as in the process in laboratory conditions. First, the outer moulds are separated, and then the inner cores are pulled out. As Neoprene has higher hardness (Shore A 35) compared to the hardness of the ELASTOSIL® VARIO mixture (Shore A 30) that we used in the laboratory conditions, it was much harder to remove the cores from the corrugated actuator. Therefore, we decided to take them out on the other side through a slightly larger opening (radius 4 mm on the top, rather than 1.75 mm on the bottom), and then those small holes were closed at the end by gluing conical plugs. The total production time for one actuator is about 1 hour. The images of the industrial process are shown in Fig. A.5.

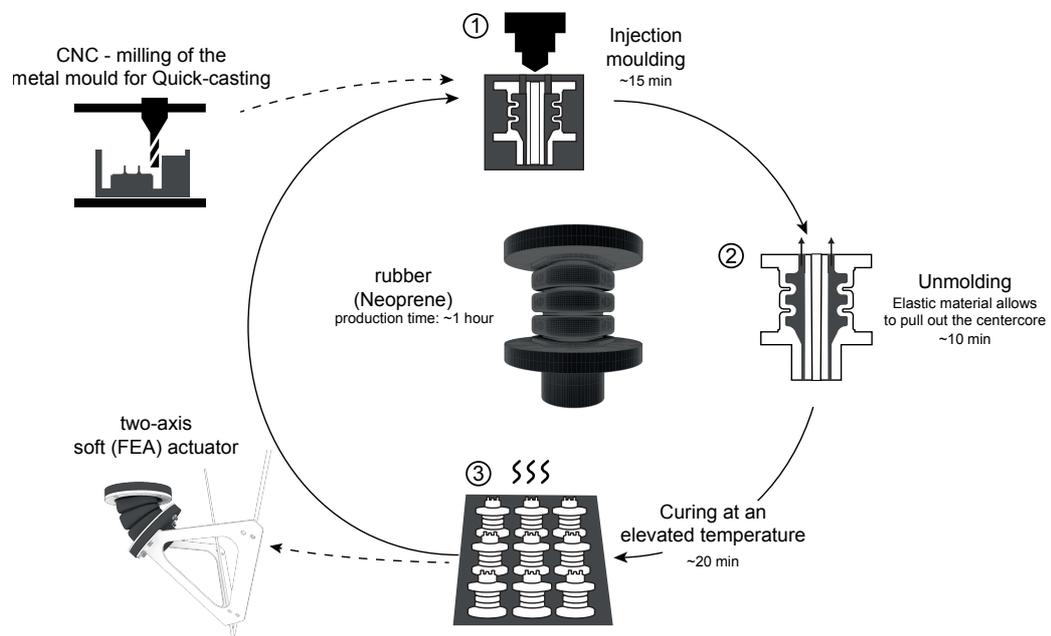

**Fig. 3.** Quick-cast manufacturing method for scalable production of FEAs in industrial setting

## 3 Results and discussion

We tested two batches of actuators with 10 actuators in each batch. The actuators are randomly selected from larger batches of about 50 actuators. The first batch consists of cylindrical SoRo-Track actuators made in laboratory from ELASTOSIL® VARIO 15/40 with the mixture ration 3:1 (resulting hardness Shore A 30). The second batch consists of corrugated actuators manufactured under industrial conditions from Neoprene rubber (hardness Shore A 35). Our goal with these tests is to compare the actuators within each batch among each other to see what degree of performance similarity can be achieved. All the tests were done using the pneumatic control system and inertial measurement unit as orientation sensor, as described in [19].

3.5 Pressure-deflection characterisation

The pressure-deflection behaviour of cylindrical actuators manufactured in laboratory conditions is shown in Fig. 4a and 4b, for roll and pitch angles, respectively. Pressure-deflection curves for each of the three chambers are presented. The measurements are done with the SoRo-Track actuator mounted at 45º roll (vertical angle), resembling the mounting angle on the façade. In that case, one chamber is below the other two chambers. Therefore, in Fig. 4a we can see that inflation of chamber 1 moves the panel vertically up, thus increasing the roll angle, while the inflations of chambers 2 and 3 cause a bending of the panel towards ground, thus decreasing the roll angle. In azimuth angle (pitch angle), the inflation of chamber 2 decreases the pitch angle, the inflation of chamber 3 increases the pitch angle, while chamber 1 does not have any influence on the pitch angle. This can be also seen on roll-pitch graph shown in Fig. 4c, where the main directions of actuation nicely follow the indicated 120º angle difference.

Regarding the differences among the actuators within the same batch, we can see that there is a significant spread in the obtained pressure- deflection curves (22%-26% difference in pressure change for 7%-23% in achieved angles). One reason for this is in manual preparation of ELASTOSIL® VARIO 15/40 mixture for each actuator before pouring. The second reason is in the insufficient precision of the 3d printed nylon mould, hence small offsets in the fixation of the inner cores during mould assembly, as well as a slight bending of the metal rods on which these cores are mounted. These metal rods are rather thin (3 mm stainless steel rod) and they bend during the manual pulling out of the inner cores (Fig. A.3).

The pressure-deflection curves for corrugated SoRo-Track actuators made from Neoprene rubber and manufactured industrially are shown in Fig. 4e-g. We can see that the pressure-deflection curves overlap for all 10 of tested actuators is very good (1% in pressure change for 12%-16% in achieved angles). As the mould is micromachined from stainless steel, there are no geometrical differences between these actuators. The only source of uncertainty is in the preparation of the rubber mix, which, due to the early development phase of establishing of this process, is still done manually for each actuator separately. However, the absolute angle error is about +/- 3º, which is acceptable for controlling the system in feed-forward mode, i.e. only using pre-calculated control inputs, without the need for feedback signals from sensors, thus simplifying the control and reducing the actuator implementation costs.

### 3.6 Repeatability of motion

One of the actuators from each batch is randomly chosen for testing the repeatability of motion (Fig. 4d and 4h). Chamber 1 was inflated and then deflated 50 times. Both actuators show very good repeatability over number of cycles.

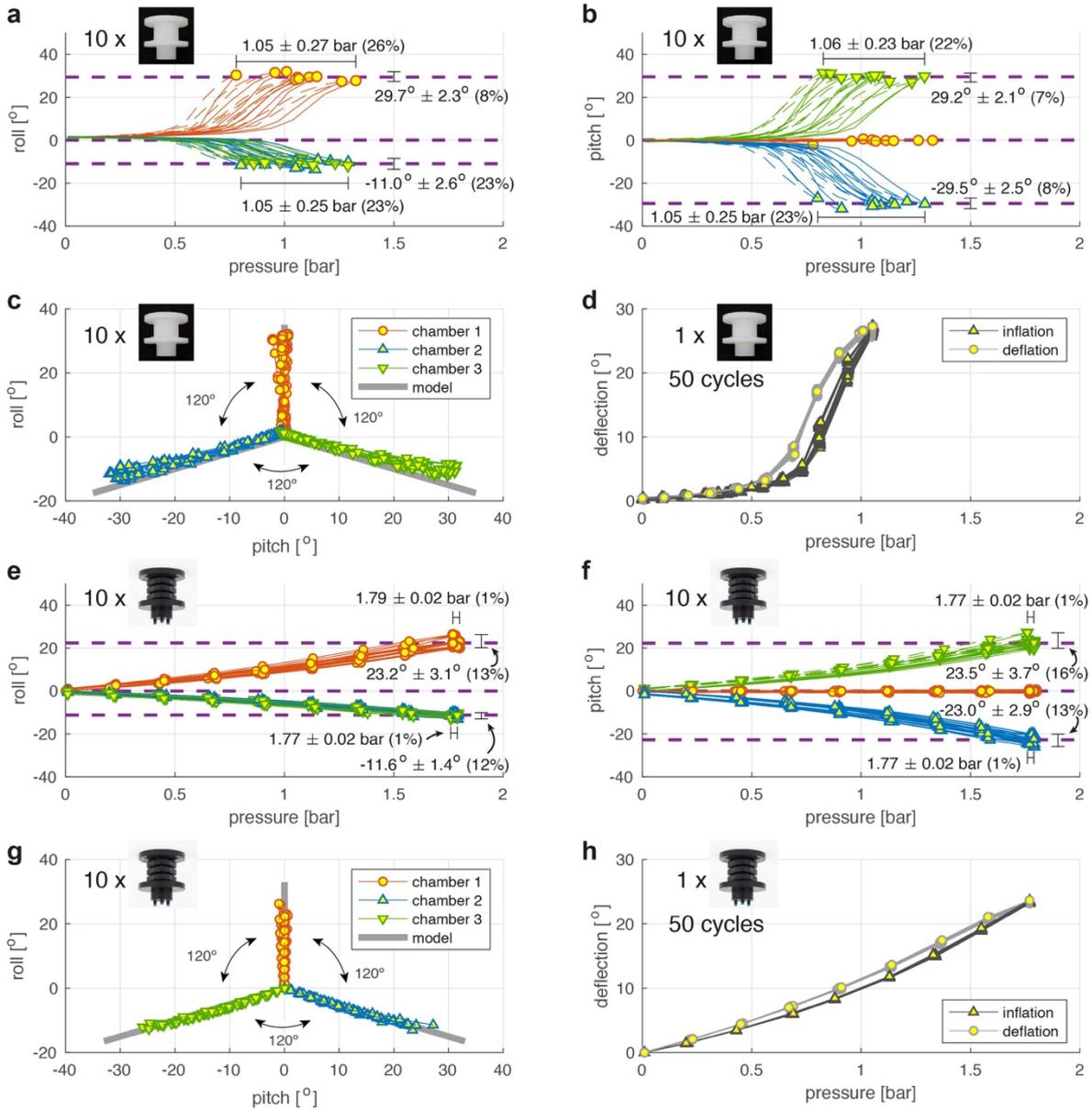

**Fig. 4.** Pressure-deflection (a-c and e-g) and repeatability characterisation (d and h) of cylindrical two-axis three-chamberd (SoRo-Track) actuator made of ELASTOSIL® VARIO in laboratory conditions and of corrugated actuator made of Neoprene in industrial conditions.

### 3.7 Comparison with state-of-the-art methods

We tested the manufacturing of SoRo-Track actuator using some the state-of-the-art methods (PneuNets [13], Wax-cores [14], and Rotation casting [25]). We did not have access to any of the direct 3d printing of soft materials techniques [21]–[23]. It turned out that manufacturing of SoRo-Track was challenging for some of the above methods, due to its specific shape, where the maximum radius of fluidic pathways is several times (5 times) larger than the radius of the fluidic pathway entrance. The actuators fabricated using PneuNets approach [13] started to delaminate after certain time. We did not have major difficulties with Wax-cores [14], but due to the softness of wax and very complex 3d geometry of SoRo-Track, we were not always sure that all parts of the wax cores stayed intact during demoulding them and afterward while pouring of elastomers over them in the main fabrication step. In terms of using the rotational casting method [25], we could not tune it for ELASTOSIL® M4601 to obtain the precise geometry of the inner cores. However, this could have been due to our lack of expertise in this method. The total time for manufacturing of a single SoRo-Track actuator using the above methods was between 6h to 10h, after basic preparations were done. We are confident in these numbers, as all of the above methods require multiple steps (between 3 and 4).

Based on our review of the state-of-the-art methods provided in Section 2.2 and based on the results of our Quick-cast method, we provide the summary of the comparison in Table 1. With the table fields shaded in grey we emphasize which methods show the best performance for each feature analysed. Overall, our method shows comparable or advanced performance across all features besides one (geometrical complexity of fluidic pathways), where it shows moderate performance. Our method shows advance performance in terms of number of steps (reduction to a single step), precision, and total fabrication time (reduction from one working day to one hour).

In terms of the geometrical complexity of fluidic pathways, our process has certain limitations defined by the materials properties and certain geometrical parameters of fluidic pathways. The difference between the radius of the entrance of the fluidic pathways and the maximum radius of the fluidic pathways at any point, should not be larger than the elongation at break of a soft material. Given that the elongation at break of soft materials is in the range of 100% to 1000%, this may not be a critical limitation. The elongation at break is inversely correlated with the elastomer hardness. This means that for stronger soft actuators (with higher elastomer hardness), the elongation at break is getting smaller. A designer of a soft actuator would need to check if the design satisfies this limitation before fabricating the actuator.

| Features | PneuNets [13] | Wax-cores [14] | Rotation casting [25] | 3d printing of soft materials [21]–[23] | Quick-cast |
|---|---|---|---|---|---|
| Leak-free actuator | (yes) Requires bonding | yes | yes | (yes) | **yes** |
| Number of steps | 3 | 4 | 3 | 1 | **1** |
| Steps complexity | low | low / moderate | moderate / high | moderate / high | **low** |
| Additional engineering effort required to tune the process for another elastomer | no | low | moderate / high | moderate / high | **no** |
| Single-body form (no bonding / gluing step) | no | yes | yes | yes | **yes** |
| Total fabrication time for SoRo-Track actuators | 6-8h | 6-8h | 10h | (est.) 12h | **1h** |
| Precision | < 1 mm | (< 1 mm) | unknown | ~ 1 mm | **< 0.2 mm** |
| Supported elastomers | wide range | wide range | limited range | unknown | **wide range** |
| Geometrical complexity of fluidic pathways | low / moderate | moderate / high | low / moderate | moderate / high | moderate |
| Based on standard industrial process | no | no | yes | no | **yes** |

**Table 1.** Comparison of Quick-cast manufacturing method with state-of-the-art methods. The fields highlighted in grey indicate the best performing method for that feature.

## 4 Conclusions

In summary, we demonstrated a new manufacturing method for scalable production of complex, industrial grade FEAs. Compared to the state-of-the-art methods, Quick-cast offers advantages in terms of: (i) speed (total fabrication time of one hour, compared to typical fabrication time of one working day, 8h, of other methods), (ii) quality of the actuator (leak-free, single-body, with < 1 mm precision), and (iii) the range of supported elastomers in terms of viscosity, pot life, and Young's modulus. We developed this process for two different settings: for laboratory conditions with 3d printing moulds for fast prototyping and using multi-component rubbers, and for industrial setting with moulds micromachined in metal and using compression moulding. We showed the application of those methods in the scalable manufacturing of the two-axis, three-chamber FEA actuator SoRo-Track. These actuators are then applied as motion drivers in adaptive solar building facades. We tested two batches of two types of actuators, with cylindrical and corrugated chamber walls, in terms of pressure-deflection and repeatability of motion. The industrial process showed advantages over the laboratory process in terms of repeatable performance over a large number of samples – the pressure-deflection curves overlap very well for 10 actuators, randomly selected from a batch of more than 50.

The only potential limitation of our approach is in the maximal achievable geometrical complexity of the fluidic pathways. The difference between the radius of the entrance of the fluidic pathways and the maximum radius of the fluidic pathways at any point, should not be larger than the elongation at break of a soft material. The elongation at break is inversely correlated with the elastomer hardness, which means that for a stronger soft actuator (with higher elastomer hardness), the elongation at break is smaller. Given that the elongation at break of soft materials is in the range of 100% to 1000%, this may not be a critical limitation.

In the paper we demonstrate the scalable production of two-axis three-chambered FEA called SoRo-Track. The short manufacturing time of the Quick-cast method here allowed for rapid scalable manufacturing of large number (>100) of these actuators and has enabled the construction of several real-world building-scale prototypes of Adaptive Solar Façades [17], [6]. The Quick-cast, however, is also suitable for other single or multi-chamber designs. The method is also suitable for larger scale actuators. We believe, that this novel manufacturing method for fast and precise scalable production of complex, high quality FEAs will foster and enable novel application of soft robotics in the future.

## 5 Acknowledgements

We acknowledge support from the Building Technologies Accelerator program of Climate-KIC. We acknowledge the support from our industrial partner Maagtechnic AG, Duebendorf, Switzerland, on establishing the industrial manufacturing process at their facilities.

**Appendix A. Supplementary data**

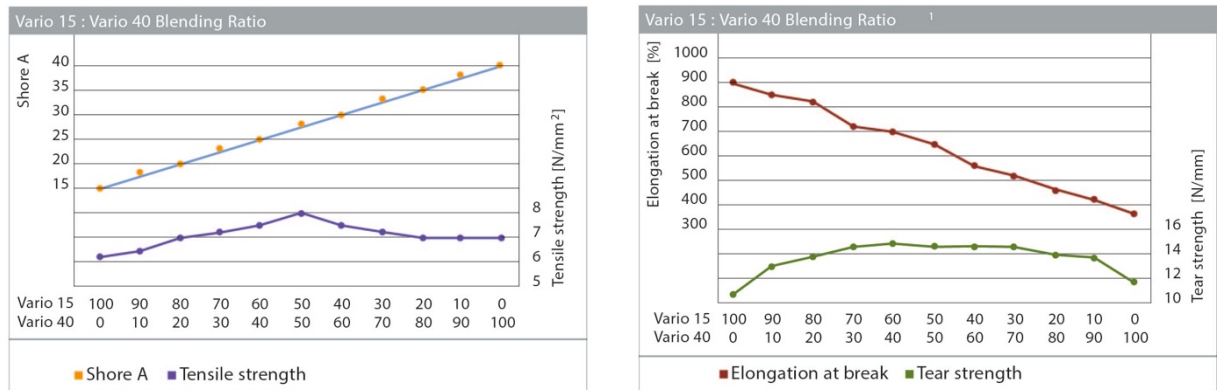

**Fig. A.1.** ELASTOSIL® VARIO is a two-component silicone rubber with variable Shoare A hardness. Influence of blending ratio of VARIO 15 : VARIO 40 on rubber parameters.

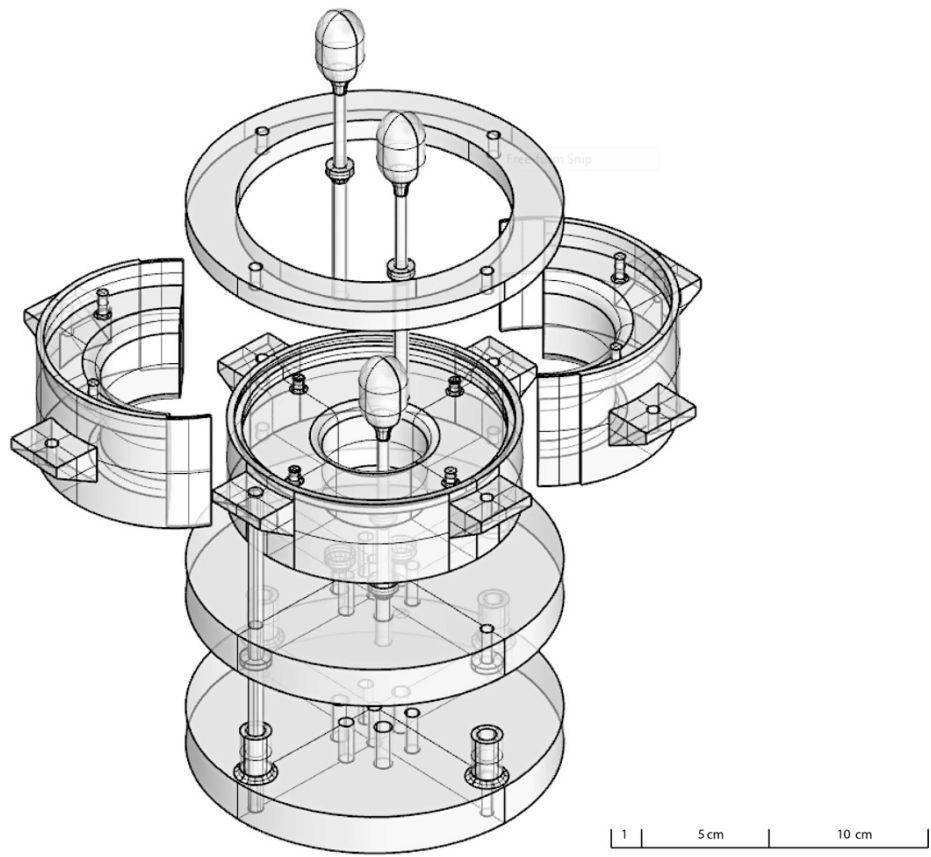

**Fig. A.2.** Exploded view of the nylon mould for the two-axis three-chambered soft actuators (SoRo-Track) with cylindrical walls. For corrugated walls, the outer moulds are the same, but the "bullets" forming the internal voids are different (see Fig. A.5).

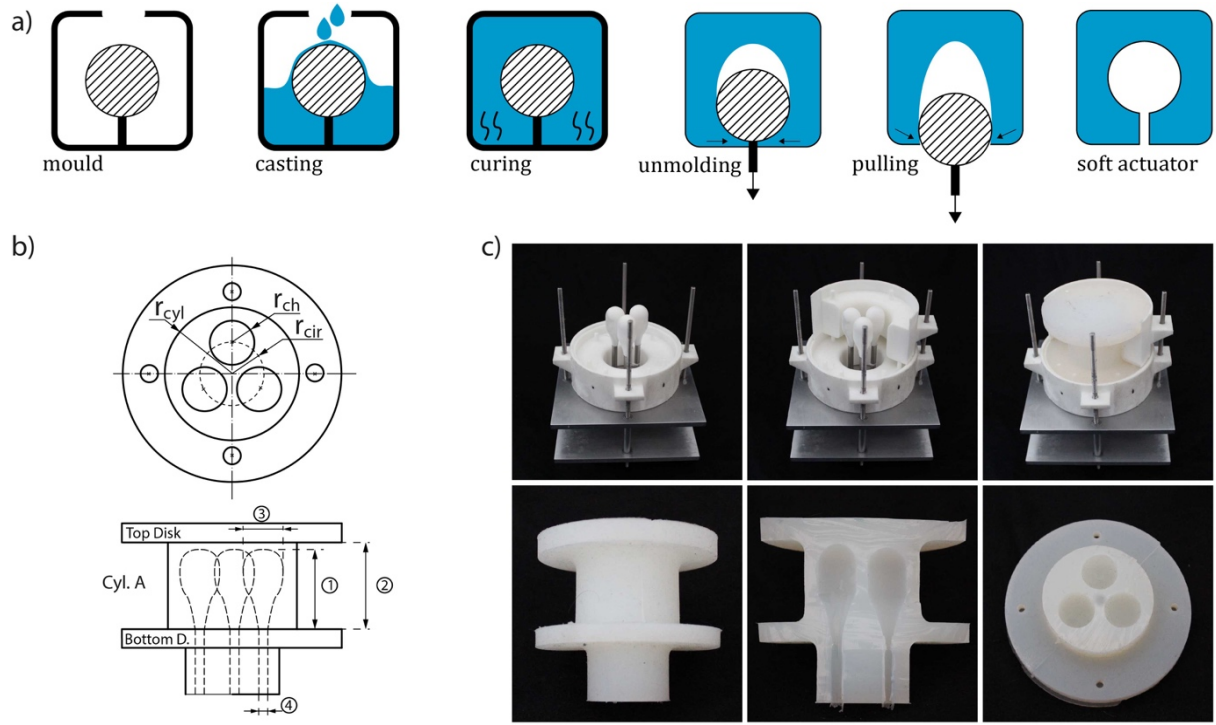

**Fig. A.3.** Manufacturing of cylindrical two-axis, three-chambered FEA actuator (SoRo-Track) in laboratory conditions with 3d printed Nylon mould from ELASTOSIL® VARIO 15/40. a) Process steps. b) Mould geometry. c) Images of fabrication steps.

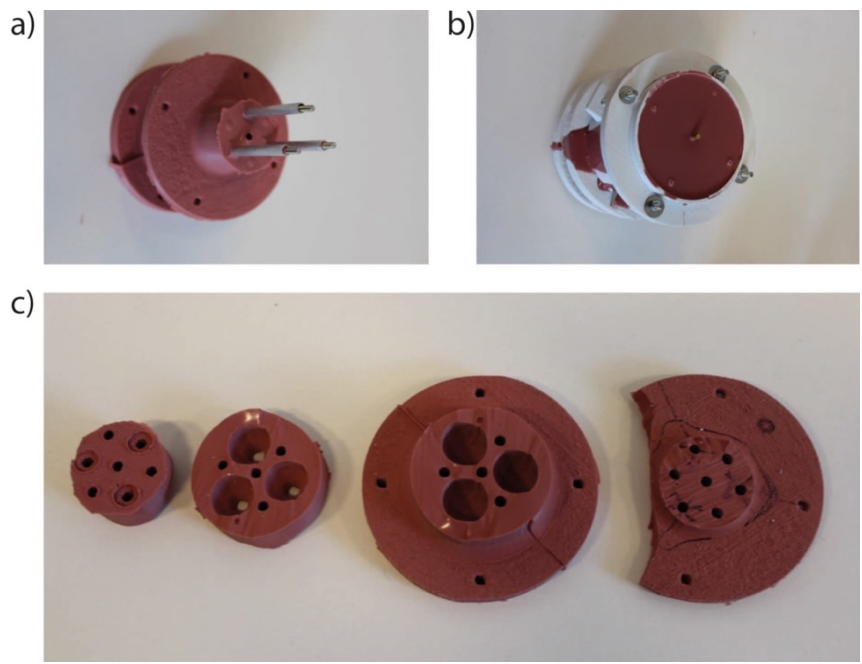

**Fig. A.4.** Manufacturing of cylindrical SoRo-Track actuator in laboratory conditions using ELASTOSIL® M4601.

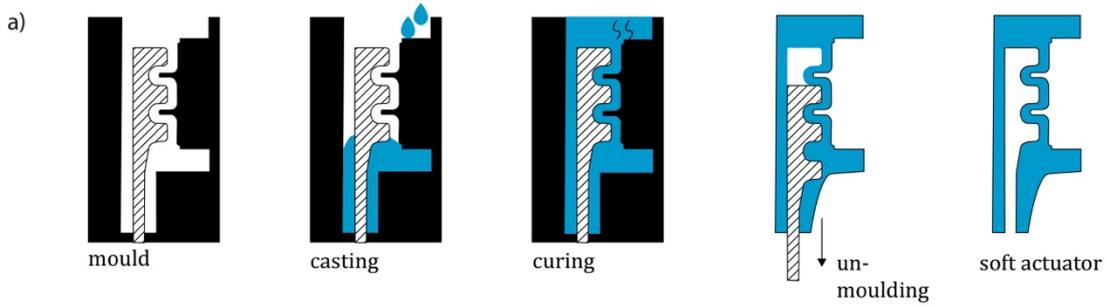
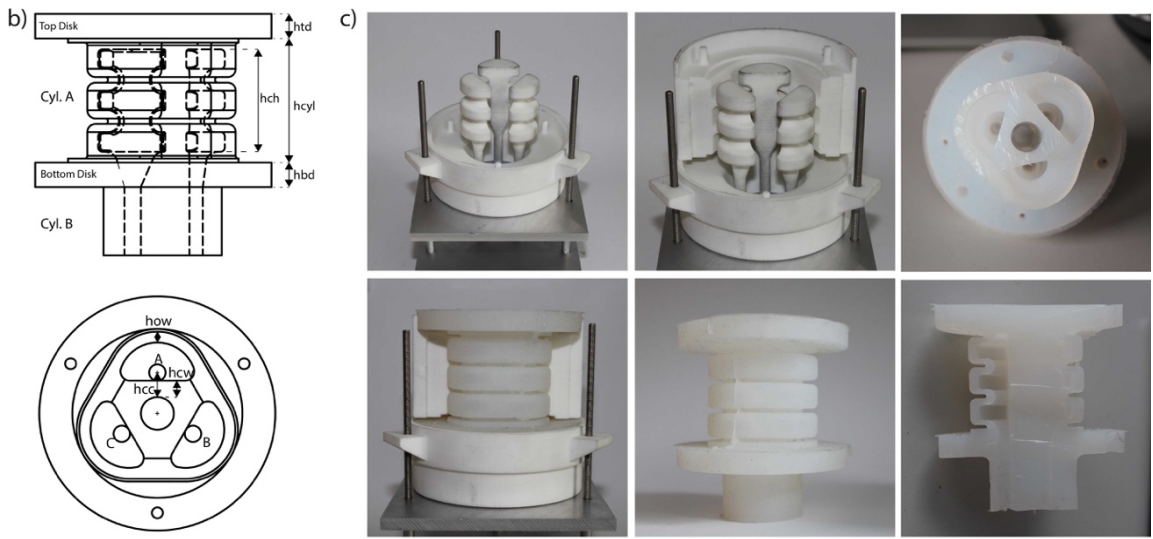

Fig. A.5. Manufacturing of corrugated two-axis three-chamebred FEA (SoRo-Track) actuator in laboratory conditions with 3d printed Nylon mould using ELASTOSIL® VARIO 15/40. a) Process steps. b) Mould geometry. c) Images of fabrication steps.

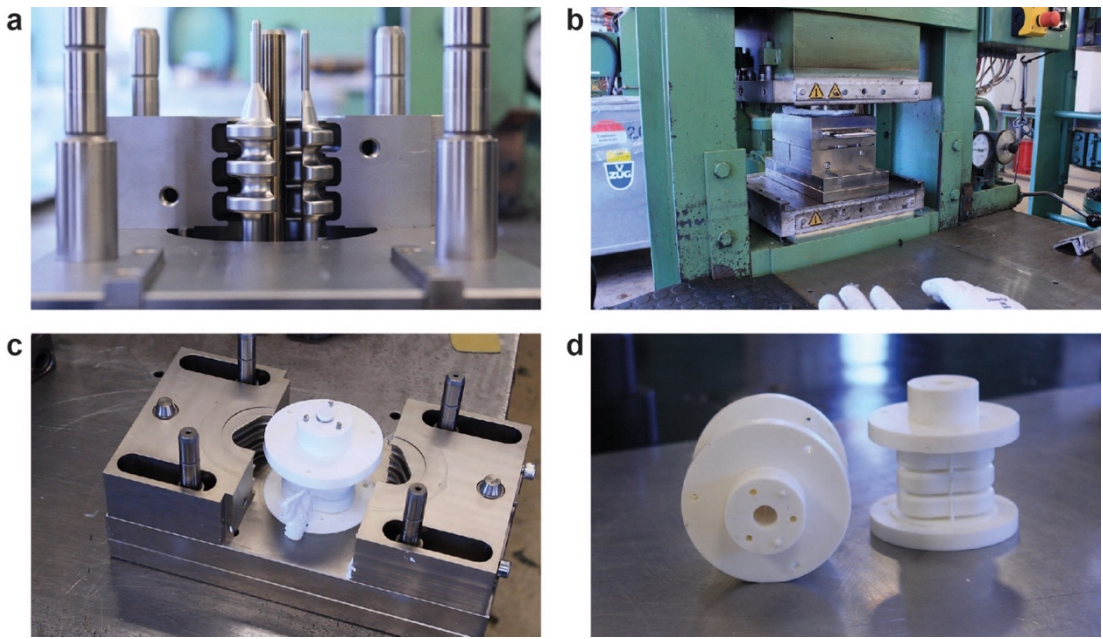

Fig. A.6. Quick-cast industrial manufacturing of corrugated two-axis three-chambered FEA (SoRo-Track) based on compression moulding and chloroprene rubber.